\documentstyle{article}
\begin{document}
\title {
\bf \huge
Boundary Kondo impurities in the generalized supersymmetric $t-J$ model
}

\author{
{\bf Heng Fan$^{a,b}$, Miki Wadati$^a$, Rui-hong Yue$^b$}\\
\normalsize
$^a$Department of Physics, Graduate School of Science,\\
\normalsize University of Tokyo, Hongo 7-3-1,\\
\normalsize Bunkyo-ku, Tokyo 113-0033, Japan.\\
\normalsize $^b$Institute of Modern Physics, P.O.Box 105,\\
\normalsize Northwest University, Xi'an 710069, P.R.China.
}

\maketitle

\begin{abstract}
We study the generalized supersymmetric $t-J$ model with Kondo
impurities in the boundaries.
We first construct the higher spin operator K-matrix for the XXZ
Heisenberg chain. Setting the boundary parameter to be a special
value, we find a higher spin reflecting K-matrix
for the supersymmetric $t-J$ model.
By using the
Quantum Inverse Scattering Method, we obtain the eigenvalue and the
corresponding Bethe ansatz equations.
\end{abstract}


\newpage
\baselineskip 0.5truecm

\section{Introduction}
There has been extensive interests in the investigation of
low-dimensional correlated electron systems with impurities.  
Recently, using renormlization group techniques,
Kane and Fisher \cite{KF} studied the transport
properties of a 1D interacting electron gas in the
presence of a potential barrier. They showed that a single
potential scatter may dramatically influence the physics in
the presence of repulsive $e-e$ interactions.
The system behaves like a Tomonaga-Luttinger liquid
rather that a Fermi liquid. Some different techniques were also
applied to study similar systems\cite{CFS, T}. The Kondo
impurities in a Tomonaga-Luttinger liquid have been investigated
in great detail \cite{AL,FJ,LT,FN}.
 
Attempts to study the effects due to the presence of impurities in
1D quantum chains in the framework of integrable models have a long
succesful history\cite{AFL}-\cite{BEF}. Andrei and Johannesson\cite{AJ}
studied an arbitrary spin S embeded in a spin-1/2 Heisenberg chain.
This method was generalized to other cases.
Recently the
supersymmetric $t-J$ model with impurities has attracted considerable
interests.
The Hamiltonian of the $t-J$ model includes the
near-neighbour hopping ($t$) and antiferromagnetic exchange ($J$)
\cite{A,ZR}
\begin{eqnarray}
H=\sum _{j=1}^L\left\{ -t{\cal {P}}
\sum _{\sigma =\pm 1}(c_{j, \sigma}^{\dagger }
c_{j+1, \sigma }+c_{j+1,\sigma }^{\dagger }c_{j,\sigma} ){\cal {P}}
+J({\bf S}_j{\bf S}_{j+1}
-{1\over 4}n_nn_{j+1})\right\}.
\end{eqnarray}
It is known that this model is supersymmetric and integrable
for $J=\pm 2t$ \cite{L,S}. The supersymmetric $t-J$ model was also
studied in Refs.\cite{S1,BB,S2,EK},
for a review, see Ref.\cite{S3} and the references
therein.
Essler and Korepin $et~al~$ \cite{EK}showed
that the one-dimensional Hamiltonian can be
obtained from the transfer matrix of
the two-dimensional supersymmetric exactly solvable lattice model
\cite{S2,FK}.

By use of the Qunatum Inverse Scattering Method (QISM) \cite{KIB},
the supersymmetric $t-J$ model with higher spin impurity was first
investigated in the periodic boundary conditions \cite{BEF}.
Recently, the supersymmetric $t-J$ model with impurities have
been studied extensively in both periodic and reflecting (open) boundary
conditions\cite{WDHP, ZGLG, FLT, FS}.

The open boundary condition was studied extensivly in the last
decade. There have been several methods to study the
the problem of open boundary condition\cite{G, ABBBQ}.    
In the end of 80's, Sklyanin\cite{S4} proposed a systematic
approach to handle the
open boundary condition problem in the framework of the QISM.
Besides the Yang-Baxter equation\cite{YB},
the reflection equation proposed by Cherednik\cite{C} also plays a
key role in proving the commutativity of the trasfer matrix.
We know that the Hamiltonian of the model is usually written as the
logarithmic derivative of a transfer matrix at zero spectral parameter.
The boundary terms in the Hamiltonina are determined by the reflecing
K-matrix which is a solution to the reflection equation. In the
usual boundary problem,
the K-matrix is a c-number matrix. The operator K-matrices which
determine the Kondo impurities in the Hamiltonian are studied
recently for several models\cite{WV},
including the supersymmetric $t-J$ model\cite{WDHP,ZGLG}.

The Hamiltonian (1) of the supersymmetric $t-J$ model can be
obtained from the transfer matrix constructed by the rational R-matrix.
We can also use the trigonometric R-matrix to formulate the transfer
matrix. The corresponding Hamiltonina is a
generalization of the original
supersymmetric $t-J$ model\cite{FK1}. This Hamiltonian satisfies a
symmetry of the quantum group $SU_q(2|1)$.
In this paper, we shall study the generalized supersymmetric $t-J$
model with higher spin boundary impurities. The operator K-matrix
is first constrcuted for the XXZ Heisenberg spin chain with
higher spin impurities. We then find a higher spin operator K-matrix
for the supersymmetric $t-J$ model. Using the graded algebraic
Bethe ansatz method, We obtain the eigenvalue of the transfer matrix and
the Bethe ansatz equations.
                                   
The paper is organized as follows: We introduce the
model in section 2. In section 3, we study the XXZ spin chain with
higher spin Kondo impurities and present the higher spin
reflecting matrices for the generalized supersymmetric $t-J$ model
are obtained. In section 4, using
the nested algebraic Bethe ansatz method, we obtain the eigenvalues of
the transfer matrix for the generalized supersymmetric $t-J$ model.
Section 5 includes a brief summary and discussions.

\section{The Model}
We first review the generalized supersymmetric $t-J$ model.
For convenience,
we choose similar notations as those in \cite{EK} and our previous
paper\cite{FW}. The Hamiltonian of the generalized
supersymmetric $t-J$ model
take the following form:
\begin{eqnarray}
H&=&\sum _{j=1}^N\sum _{\sigma =\pm }
[c_{j,\sigma }^{\dagger}(1-n_{j,-\sigma })c_{j+1,\sigma }
(1-n_{j+1,-\sigma})
+c_{j+1,\sigma }^{\dagger}(1-n_{j+1,-\sigma })c_{j+1,\sigma }
(1-n_{j,-\sigma})]
\nonumber \\
&&-2\sum _{j=1}^N[{1\over 2}(S_j^{\dagger }S_{j+1}+S_jS_{j+1}^{\dagger })
+cos(\eta )S_j^zS_{j+1}^z
-{cos(\eta )\over 4}n_jn_{j+1}]
\nonumber \\
&&+isin(\eta )\sum _{j=1}^N[S_j^zn_{j+1}-S_{j+1}^zn_j].
\end{eqnarray}
When the anisotropic parameter $\eta =0$, this Hamiltonin
reduces to an equivalent form of the original Hamiltonian (1).
The operators $c_{j,\sigma }$ and $c_{j,\sigma }^{\dagger }$ mean the
annihilation and creation operators of electron with spin $\sigma $
on a lattice site $j$, and we assume the total number of
lattice sites is $N$, $\sigma =\pm $ represent
spin down and up, respectively. These operators are canonical
Fermi operators satisfying anticommutation relations
\begin{eqnarray}
\{c_{j,\sigma }^{\dagger },c_{j,\tau }\} =\delta _{ij}
\delta _{\sigma \tau }.
\end{eqnarray}
We denote by $n_{j,\sigma }=c_{j,\sigma }^{\dagger }c_{j,\sigma }$
the number operator for the electron on a site $j$ with
spin $\sigma $, and by
$n_j=\sum _{\sigma =\pm }n_{j,\sigma }$ the number operator for the
electron on a site $j$.
The Fock vacuum state $|0>$ is defined as $c_{j,\sigma }|0>=0$.
Due to the exclusion of double occupancy,
there are
altogether three possible electronic states at a given lattice site
$j$
\begin{eqnarray}
|0>,~~~|\uparrow >_j=c_{j,1}^{\dagger }|0>,~~~
|\downarrow >_j=c_{j,-1}^{\dagger }|0>.
\end{eqnarray}
$S^z_j,S_j,S_j^{\dagger }$ are spin operators satisfying $su(2)$
algebra and can be expressed as:
\begin{eqnarray}
S_j=c_{j,1}^{\dagger }c_{j,-1},
~~~~S_j^{\dagger }=c_{j,-1}^{\dagger }c_{j,},
~~~~S_j^z={1\over 2}(n_{j,1}-n_{j,-1}).
\label{S}
\end{eqnarray}

The above Hamiltonian can be obtained from the logarithmic derivative
of the transfer matrix at
zero spectral parameter.
In the framework of QISM, the transfer matrix is constructed by
the trigonometric R-matrix of the Perk-Schultz
model \cite{PS}. The non-zero
entries of the R-matrix are given by
\begin{eqnarray}
&&{\tilde {R}}(\lambda )^{aa}_{aa}=sin(\eta +\epsilon _a\lambda ),
\nonumber \\
&&{\tilde {R}}(\lambda )^{ab}_{ab}=(-1)^{\epsilon _a\epsilon _b}
sin(\lambda ),
\nonumber \\
&&{\tilde {R}}(\lambda )^{ab}_{ba}=e^{isign(a-b)\lambda }sin(\eta ),
a\not= b,
\end{eqnarray}
where
$\epsilon _a$ is the Grassman parity,
$\epsilon _a=0$ for boson and $\epsilon _a=1$ for fermion, and
\begin{eqnarray}
sign(a-b)=\left\{
\begin{array}{ll}1,&{\rm if} ~a>b \\
-1, &{\rm if}~ a<b.
\end{array}\right.
\end{eqnarray}
This R-matrix of the Perk-Schultz model
satisfies the usual Yang-Baxter equation:
\begin{eqnarray}
{\tilde {R}}_{12}(\lambda -\mu )
{\tilde {R}}_{13}(\lambda )
{\tilde {R}}_{23}(\mu )
={\tilde {R}}_{23}(\mu )
{\tilde {R}}_{13}(\lambda )                                    
{\tilde {R}}_{12}(\lambda -\mu )
\end{eqnarray}
In this paper, we shall concentrate our discussion only to
the Fermionic, Fermionic and Bosonic case (FFB), that means
$\epsilon _1=\epsilon _2=1, \epsilon _3=0$. And we shall use
the graded formulae to study this model. For supersymmetric
$t-J$ model, the spin of the electrons and the charge `hole'
degrees of freedom play a very similar role forming a
graded superalgebra with two fermions and one boson. The holes
obey boson commutation relations, while the spinons are
fermions\cite{S3}. The graded approach has an advantage
of making clear distinction between bosonic and fermionic degrees
of freedom \cite{OWA}.

Introducing a diagonal matrix
$\Pi _{ac}^{bd}=(-)^{\epsilon _a\epsilon _c}
\delta _{ab}\delta _{cd}$, we change the original R-matrix to the following
form
\begin{eqnarray}
R(\lambda )=\Pi {\tilde {R}}(\lambda ).
\end{eqnarray}
From the non-zero elements of the R-matrix
$R_{ab}^{cd}$, we see that
$\epsilon _a+\epsilon _b+\epsilon _c+\epsilon _d=0$.
One can show that the R-matrix satisfies the graded Yang-Baxter equation
\begin{eqnarray}
R(\lambda -\mu )_{a_1a_2}^{b_1b_2}
R(\lambda )_{b_1a_3}^{c_1b_3}
R(\mu )_{b_2b_3}^{c_2c_3}
(-)^{(\epsilon _{b_1}+\epsilon _{c_1})\epsilon _{b_2}}
=
R(\mu )_{a_2a_3}^{b_2b_3}R(\lambda )_{a_1b_3}^{b_1c_3}
R(\lambda -\mu )_{b_1b_2}^{c_1c_2}(-)^{(\epsilon _{a_1}
+\epsilon _{b_1})\epsilon _{b_2}}.
\end{eqnarray}
Explicitly the R-matrix is written as
\begin{eqnarray}
R(\lambda )=\left(
\begin{array}{ccccccccc}
a(\lambda )&0&0&0&0&0&0&0&0\\                                        
0&b(\lambda )&0&-c_-(\lambda )&0&0&0&0&0\\
0&0&b(\lambda )&0&0&0&c_-(\lambda )&0&0\\
0&-c_+(\lambda )&0&b(\lambda )&0&0&0&0&0\\
0&0&0&0&a(\lambda )&0&0&0&0\\
0&0&0&0&0&b(\lambda )&0&c_-(\lambda )&0\\
0&0&c_+(\lambda )&0&0&0&b(\lambda )&0&0\\
0&0&0&0&0&c_+(\lambda )&0&b(\lambda )&0\\
0&0&0&0&0&0&0&0&w(\lambda )
\end{array}\right),
\label{R}
\end{eqnarray}
where
\begin{eqnarray}
a(\lambda )=sin(\lambda -\eta ),
~w(\lambda )=sin(\lambda +\eta ), 
~b(\lambda )=sin(\lambda ),
~c_{\pm }(\lambda )=e^{\pm i\lambda }sin(\eta ).
\end{eqnarray}
In the framework of the QISM, we can construct
the $L$ operator from the R-matrix as:
\begin{eqnarray}
L_n(\lambda )=\left(
\begin{array}{ccc}
b(\lambda )-(b(\lambda )-a(\lambda ))e^n_{11}&
-c_-(\lambda )e^n_{21} &c_-(\lambda )e^n_{31}\\
-c_+(\lambda )e^n_{12} &b(\lambda )-(b(\lambda )-a(\lambda ))e^n_{22}&
c_-(\lambda )e^n_{32}\\
c_+(\lambda )e^n_{13} &c_+(\lambda )e^n_{23} &
b(\lambda )-(b(\lambda )-w(\lambda ))e^n_{33}
\end{array}\right).                                         
\end{eqnarray}
Here $e^n_{ab}$ acts on the $n$-th quantum space.
Thus we have the (graded) Yang-Baxter relation
\begin{eqnarray}
R_{12}(\lambda -\mu )L_1(\lambda )L_2(\mu )
=L_2(\mu )L_1(\lambda )R_{12}(\lambda -\mu ).
\label{YBR}
\end{eqnarray}
Here the tensor product is in the sense of super tensor product
defined as
\begin{eqnarray}
(F\otimes G)_{ac}^{bd}=F_a^bG_c^d(-)^{(\epsilon _a+\epsilon _b)\epsilon _c}.
\end{eqnarray}
Except in Section 3.1, all tensor products in this paper
are in the super sense.

The row-to-row monodromy matrix
$T_N(\lambda )$ is defined as a matrix product over
the $N$ operators on all sites of the lattice, 
\begin{eqnarray}
T_a(\lambda )=L_{aN}(\lambda )L_{aN-1}(\lambda )\cdots L_{a1}(\lambda ), 
\end{eqnarray}
where the subscript $a$ represents the auxiliary space, and
the tensor product is in the graded sense. Explicitely we write
\begin{eqnarray}
&&\{ [T(\lambda )]^{ab}\}_{\begin {array}{c}
\alpha _1\cdots \alpha _N\\
\beta _1\cdots \beta _N\end{array}}
\nonumber \\
&=&L_N(\lambda )_{a\alpha _N}^{c_N\beta _N}
L_{N-1}(\lambda )_{c_N\alpha _{N-1}}^{c_{N-1}\beta _{N-1}}
\cdots L_1(\lambda )_{c_2\alpha _1}^{b\beta _1}
(-1)^{\sum _{j=2}^N(\epsilon _{\alpha _j}+\epsilon _{\beta _j})
\sum _{i=1}^{j-1}\epsilon _{\alpha _i}}
\end{eqnarray}
By repeatedly using the Yang-Baxter relation (\ref{YBR}),
one can prove easily that
the monodromy matrix also satisfies the Yang-Baxter relation
\begin{eqnarray}
R(\lambda -\mu )T_1(\lambda )T_2(\mu )
=T_2(\mu )T_1(\lambda )R(\lambda -\mu ).
\label{YBR1}
\end{eqnarray}
For periodic boundary condition, the transfer
matrix $\tau _{peri}(\lambda )$
of this model is defined as the supertrace of the
monodromy matrix in the auxiliary space
\begin{eqnarray}
\tau _{peri}(\lambda )=strT(\lambda )
=\sum (-1)^{\epsilon _a}T(\lambda )_{aa}.
\end{eqnarray}
As a consequence of the Yang-Baxter relation (\ref{YBR1}) and the unitarity
property of the R-matrix, we can prove that
the transfer matrix commutes with each other for different spectral
parameters,
\begin{eqnarray}
[\tau _{peri}(\lambda ),\tau _{peri}(\mu )]=0. 
\end{eqnarray}
In this sense we say that the model is
integrable. Expanding the transfer matrix in the powers of $\lambda $,
we can find conserved quantites. The first non-trivial conserved
equantity is the Hamiltonian (1),
\begin{eqnarray}
H=sin(\eta )
\frac {d\ln [\tau (\lambda )]}{d\lambda }|_{\lambda =0}
=\sum _{j=1}^NH_{j,j+1}=
\sum _{j=1}^NP_{j,j+1}L'_{j,j+1}(0),
\end{eqnarray}
where $P_{ij}$ is the graded permutation operator expressed as
$P_{ac}^{bd}=\delta _{ad}\delta _{bc}(-1)^{\epsilon _a\epsilon _c}$.

In this paper, we consider the reflecting boundary condition case.
In addition to the Yang-Baxter equation, a reflection equation should
be used in proving the commutativity of the transfer matrix with
boundaries. The reflection equation takes the form \cite{C}
\begin{eqnarray}
R_{12}(\lambda -\mu )K_1(\lambda )R_{21}(\lambda +\mu )
K_2(\mu )=K_2(\mu )R_{12}(\lambda +\mu )K_1(\lambda )         
R_{21}(\lambda -\mu ).
\end{eqnarray}
For the graded case, the reflection equation remains the
same as the above form.
We only need to change the usual tensor product to the graded
tensor product. We write it explicitly as
\begin{eqnarray}
&&R(\lambda -\mu )_{a_1a_2}^{b_1b_2}K(\lambda )_{b_1}^{c_1}
R(\lambda +\mu )_{b_2c_1}^{c_2d_1}                           
K(\mu )_{c_2}^{d_2}(-)^{(\epsilon _{b_1}+\epsilon _{c_1})\epsilon _{b_2}}
\nonumber \\
&=&K(\mu )_{a_2}^{b_2}R(\lambda +\mu )_{a_1b_2}^{b_1c_2}     
K(\lambda )_{b_1}^{c_1}
R(\lambda -\mu )_{c_2c_1}^{d_2d_1}                            
(-)^{(\epsilon _{b_1}+\epsilon _{c_1})\epsilon _{c_2}}.
\label{RK}
\end{eqnarray}
Instead of the monodromy matrix $T(\lambda )$ for periodic boundary
conditions, we consider the double-row monodromy matrix
\begin{eqnarray}
{\cal {T}}(\lambda )=T(\lambda )K(\lambda )T^{-1}(-\lambda )
\end{eqnarray}
for the reflecting boundary conditions. Using the Yang-Baxter relation, and
considering the boundary K-matrix which satisfies
the reflection equation, one
can prove that the double-row monodromy matrix ${\cal {T}}(\lambda )$
also satisfies the reflection equation
\begin{eqnarray}
&&R(\lambda -\mu )_{a_1a_2}^{b_1b_2}{\cal {T}}(\lambda )_{b_1}^{c_1}
R(\lambda +\mu )_{b_2c_1}^{c_2d_1}                           
{\cal {T}}(\mu )_{c_2}^{d_2}
(-)^{(\epsilon _{b_1}+\epsilon _{c_1})\epsilon _{b_2}}
\nonumber \\
&=&{\cal {T}}(\mu )_{a_2}^{b_2}R(\lambda +\mu )_{a_1b_2}^{b_1c_2}     
{\cal {T}}(\lambda )_{b_1}^{c_1}
R(\lambda -\mu )_{c_2c_1}^{d_2d_1}                            
(-)^{(\epsilon _{b_1}+\epsilon _{c_1})\epsilon _{c_2}}.
\label{RT}
\end{eqnarray}

Next, we study the properties of the R-matrix.
We define the super-transposition $st$ as
\begin{eqnarray}
(A^{st})_{ij}=A_{ji}(-1)^{(\epsilon _i+1)\epsilon _j}.
\end{eqnarray}
For FFB grading used in this paper, $\epsilon _1=\epsilon _2=1,
~\epsilon _3=0$,
we can rewrite the above relation explicitly as
\begin{eqnarray}
\left( \begin{array}{ccc}
A_{11}&A_{12}&B_1\\
A_{21}&A_{22}&B_2\\
C_1&C_2&D\end{array}\right)^{st}                            
=\left( \begin{array}{ccc}
A_{11}&A_{21}&C_1\\
A_{12}&A_{22}&C_2\\
-B_1&-B_2&D\end{array}\right).
\end{eqnarray}
We also define the inverse of 
the super-transposition $\bar {st}$ as 
$\{ A^{st}\} ^{\bar {st}}=A$.

One can prove directly that the
R-matrix (\ref{R}) satisfy the following unitarity and cross-unitarity
relations:
\begin{eqnarray}
&&R_{12}(\lambda )R_{21}(-\lambda )=\rho (\lambda )\cdot id.,~~~~
\rho (\lambda )=sin(\eta +\lambda )sin(\eta -\lambda ),    \\
&&R_{12}^{st_1}(\eta -\lambda )M_1
R_{21}^{st_1}(\lambda )M_1^{-1}={\tilde {\rho }}(\lambda )\cdot id.,~~~~
\tilde {\rho }(\lambda )=sin(\lambda )sin(\eta -\lambda ).
\end{eqnarray}
Here the matrix $M=diag.(e^{2i\eta },1,1)$ is determined by the R-matrix.
The cross-unitarity relation can also be written as the following form
\begin{eqnarray}
\left\{ M_1^{-1}R_{12}^{st_1st_2}(\eta -\lambda )M_1
\right\} ^{\bar {st}_2}R_{21}^{st_1}(\lambda )&=&\tilde {\rho }(\lambda ),\\
R_{12}^{st_1}(\lambda )\left\{ M_1R_{21}^{st_1st_2}(\eta -\lambda )
M_1^{-1}\right\} ^{\bar {st}_2}&=&\tilde {\rho }(\lambda ).
\end{eqnarray}
In order to construct the commuting transfer matrix with boundaries,
besides the
reflection equation, we need the dual reflection equation.
In general, the dual reflection equation which depends on
the unitarity and cross-unitrarity relations of the R-matrix   
takes different forms for different models. 
For the models considered in this paper,
we can write the dual
reflection equation in the following form:
\begin{eqnarray}
&&R_{21}^{st_1st_2}(\mu -\lambda )
{K_1^+}^{st_1}(\lambda )M_1^{-1}R_{12}^{st_1st_2}(\eta -\lambda -\mu )
M_1{K_2^+}^{st_2}(\mu )     
\nonumber \\
&=&{K_2^+}^{st_2}(\mu )M_1R_{21}^{st_1st_2}(\eta -\lambda -\mu )
M_1^{-1}K_1^+(\lambda )R_{12}^{st_1st_2}(\mu -\lambda ).
\label{DRK}
\end{eqnarray}
Then the transfer matrix with boundaries is defined as:
\begin{eqnarray}
t(\lambda )=strK^+(\lambda ){\cal {T}}(\lambda ).
\label{tran}
\end{eqnarray}
The commutativity of $t(\lambda )$ can be proved by using
unitarity and cross-unitarity relations, reflection equation and 
the dual reflection equation. The detailed proof of the commuting
transfer matrix with boundaries for super (graded)
case can be found, for instance, in Ref.\cite{BGZZ,GZZ,FHS} etc..
With a normalization $K(0)=id.$, the Hamiltonian can be obtained as
\begin{eqnarray}
H&\equiv &{1\over 2}sin(\eta )\frac {d\ln t(\lambda )}
{d\lambda}|_{\lambda =0}
\nonumber \\
&=&\sum _{j=1}^{N-1}P_{j,j+1}L'_{j,j+1}(0)+
{1\over 2}sin(\eta )K'_1(0)
+\frac {str_aK_a^+(0)P_{Na}L_{Na}'(0)}{str_{a}K_a^+(0)}.
\end{eqnarray}

\section{Higher spin solution to the reflection equation for
supersymmetric $t-J$ model}
In order to find the higher spin solution to the reflection equation
for the generalized supersymmetric $t-J$ model, we first construct
the higher spin reflecting matrix for the XXZ Heisenberg chain.

\subsection{XXZ Heisenberg chain with higher spin boundary impurities}
The higher spin R-matrix can be constructed by using the fusion procedure
\cite{KRS}.
The Hamiltonian of the XXZ Heisenberg chain is written as
\begin{eqnarray}
H_{XXZ}=\sum _{j=1}^N[\sigma _j^+\sigma _{j+1}^-
+\sigma _j^-\sigma _{j+1}^++{1\over 2}cos(\eta )\sigma _j^z\sigma _{j+1}^z].
\end{eqnarray}
Here $\sigma ^{\pm }={1/2}(\sigma ^x\pm \sigma ^y)$ and
$\sigma ^x, \sigma ^y$ and $\sigma ^z$ are Pauli matrices.
The R-matrix is known to be the standard six-vertex model,
\begin{eqnarray}
r_{12}(\lambda )=\left( \begin{array}{cccc}
sin(\lambda +\eta )&0&0&0\\
0&sin(\lambda )&sin(\eta )&0\\              
0&sin(\eta )&sin(\lambda )&0\\
0&0&0&sin(\lambda +\eta )\end{array}\right).
\label{6r}
\end{eqnarray}
In the framework of QISM, the L-operator constructed by the r-matrix
is written as:
\begin{eqnarray}
L_{ak}(\lambda )=\left( \begin{array}{cc}
sin(\lambda +{1\over 2}\eta +{1\over 2}\eta \sigma _k^z)
&sin(\eta )\sigma _k^-\\
sin(\eta )\sigma _k^+&sin(\lambda +{1\over 2}\eta -{1\over 2}
\eta \sigma _k^z)
\end{array}\right),
\end{eqnarray}
where $a$ represents auxiliary space. As usual, we can construct
the row-to-row monodromy matrix $T_a(\lambda )=L_{aN}(\lambda )
\cdots L_{a1}(\lambda )$, and we have the Yang-Baxter relation
\begin{eqnarray}
r_{12}(\lambda -\mu )T_1(\lambda )T_2(\mu )
=T_2(\mu )T_1(\lambda )r_{12}(\lambda -\mu ),
\end{eqnarray}
where the tensor product is a non-graded one.

Next, we consider the higher spin operators.
Let the higher spin L operator take the form \cite{KRS,JZ}
\begin{eqnarray}
{\cal L}(\lambda )=\left( \begin{array}{cc}
sin(\lambda +{\bf S^z}\eta ) &sin(\eta ){\bf S^-}\\
sin(\eta ){\bf S^+}&sin(\lambda -{\bf S^z}\eta )
\end{array}\right) ,
\end{eqnarray}
where ${\bf S^z, S}$ and ${\bf S^{\dagger }}$ are
spin-$s$ operators satisfying the following commutation relations,
\begin{eqnarray}                                           
[{\bf S^z}, {\bf S^{\pm }}]=\pm {\bf S^{\pm }},
~~~~~~~[{\bf S^+,S^-}]=\frac {sin(2{\bf S^z}\eta )}{sin(\eta )}.
\end{eqnarray}
We also have the following relations for spin-$s$ operator:
\begin{eqnarray}
&&sin({\bf S^z}\eta )sin(\eta +{\bf S^z}\eta )
+sin^2(\eta ){\bf S^-S^+}
\nonumber \\
&=&sin^2(\eta ){\bf S^+S^-}+sin({\bf S^z}\eta )
sin({\bf S^z}\eta -\eta )
=sin(s\eta )sin(s\eta +\eta )
\end{eqnarray}
A more general relations can be written as
\begin{eqnarray}
&&sin({\lambda +\bf S^z}\eta )sin(\eta +{\bf S^z}\eta -\lambda )
+sin^2(\eta ){\bf S^-S^+}
\nonumber \\                                                
&=&sin^2(\eta ){\bf S^+S^-}+sin(\lambda -{\bf S^z}\eta )
sin(-\lambda -{\bf S^z}\eta +\eta )
=sin(\lambda +s\eta )sin(s\eta +\eta -\lambda ).
\end{eqnarray}
One can prove that the higher spin L operator also satisfy the
Yang-Baxter relation
\begin{eqnarray}
r_{12}(\lambda -\mu ){\cal L}_1(\lambda ){\cal L}_2(\mu )
={\cal L}_2(\mu ){\cal L}_1(\lambda )r_{12}(\lambda -\mu ).
\end{eqnarray}                                             

Now, Let us consider the reflecting boundary condition. We can find
a c-number solution to the reflection equation
$K_c(\lambda )=diag.\left( sin(\xi +\lambda ), sin(\xi -\lambda )\right)$,
where $\xi $ is an arbitrary parameter.
This is a general c-number diagonal solution to the reflection equation.
In particular, if $\xi \rightarrow -i\infty $, we find
$K(\lambda )=diag.(e^{2i\lambda },1)$ is a solution to the reflection
equation.

It is interesting to find a higher spin operator
K-matrix. We can construct the operator K-matrix by
${\cal K}_{XXZ}(\lambda )={\cal L}(\lambda +c)
K_c(\lambda ){\cal L}^{-1}(-\lambda +c)$, one
can find easily that ${\cal K}(\lambda )$ is an operator reflecting matrix
satisfying the reflection equation. Explicilty, the higher spin
reflecting ${\cal K}$ has the form
${\cal K}_{XXZ}(\lambda )
=\left( \begin{array}{cc}
K(\lambda )_1^1&K(\lambda )_1^2\\
K(\lambda )_2^1&K(\lambda )_2^2\end{array}\right)$ with
\begin{eqnarray}
K(\lambda )_1^1&=&
sin(\lambda -\xi )sin(\lambda +c+s\eta )sin(\lambda +c-\eta -s\eta )
\nonumber \\
&&+sin(2\lambda )sin(\lambda +c+{\bf S^z}\eta )
sin(\xi -c+\eta +{\bf S^z}\eta ),\nonumber \\
K(\lambda )_2^2&=&-sin(\xi +\lambda )sin(\lambda +c+s\eta )
sin(\lambda +c-\eta -s\eta )
\nonumber \\
&&+sin(2\lambda )sin(\lambda +c-{\bf S^z}\eta )                    
sin(\xi +c-\eta +{\bf S^z}\eta ),\nonumber \\
K(\lambda )_1^2&=&sin(\eta )sin(2\lambda )
sin(\xi +c+{\bf S^z}\eta ){\bf S^-},
\nonumber \\
K(\lambda )_2^1&=&sin(\eta )sin(2\lambda )
sin(\xi -c+{\bf S^z}\eta ){\bf S^+}.
\end{eqnarray}
By use of the cross-unitarity relation of the r-matrix, the operator
reflecting matrix to the dual reflection equation can also be found.
The eigenvalues of the transfer matrix can be obtained by applying the
algebraic Bethe ansatz method. These results will be presented in
a seperate paper \cite{F}.

\subsection{Higher spin reflecting matrix for the supersymmetric
$t-J$ model}
We know that the generalized
supersymmetric $t-J$ model has a $SU_q(2)$ symmetry.
We suppose that the operator
K-matrix takes the following form:
\begin{eqnarray}
K(\lambda )=\left( \begin{array}{ccc}
A(\lambda ) &B(\lambda ) &0\\
B(\lambda ) &C(\lambda ) &0\\
0&0&1\end{array}\right)
\label{K}
\end{eqnarray}
Inserting this matrix into the reflection equation (\ref{RK}), we can find
the following non-trivial relations:
\begin{eqnarray}
\hat {r}(\lambda -\mu )_{a_1a_2}^{b_1b_2}K(\lambda )_{b_1}^{c_1}
\hat {r}(\lambda +\mu )_{b_2c_1}^{c_2d_1}                           
K(\mu )_{c_2}^{d_2}
&=&K(\mu )_{a_2}^{b_2}\hat {r}(\lambda +\mu )_{a_1b_2}^{b_1c_2}     
K(\lambda )_{b_1}^{c_1}
\hat {r}(\lambda -\mu )_{c_2c_1}^{d_2d_1},
\label{rk}
\end{eqnarray}
and
\begin{eqnarray}
K(\lambda )_{a_1}^{b_1}K(\mu )_{b_1}^{d_1}=
K(\mu )_{a_1}^{b_1}K(\lambda )_{b_1}^{d_1},
\label{rk1}
\end{eqnarray}
\begin{eqnarray}
\delta _{a_1d_1}sin(\lambda -\mu )e^{-i(\lambda +\mu )}
+sin(\lambda +\mu )e^{i(\lambda -\mu )}K(\lambda )_{a_1}^{d_1}
\nonumber \\
=e^{-i(\lambda -\mu )}sin(\lambda +\mu )K(\mu )_{a_1}^{d_1}  
+e^{i(\lambda +\mu )}K(\mu )_{a_1}^{b_1}K(\lambda )_{b_1}^{d_1},
\label{rk2}
\end{eqnarray}
where all indices take values 1,2, and we have introduced
\begin{eqnarray}
\hat {r}_{12}(\lambda )=\left( \begin{array}{cccc}
sin(\lambda -\eta )&0&0&0\\
0&sin(\lambda )&-sin(\eta )e^{-i\lambda }&0\\
0&-sin(\eta )e^{i\lambda }&sin(\lambda )&0\\                 
0&0&0&sin(\lambda -\eta )\end{array}\right).
\label{r}
\end{eqnarray}
This matrix $\hat {r}(\lambda )$ can be obtained from the
matrix (\ref{6r}) by a gauge
transformation and with a change $\eta \rightarrow -\eta $.
Correspondingly, we can show that 
$A(\lambda )=f(\lambda )e^{-2i\lambda }K(\lambda )_1^1$,
$B(\lambda )=f(\lambda )e^{-i\lambda }K(\lambda )_1^2$,
$C(\lambda )=f(\lambda )e^{-i\lambda }K(\lambda )_2^1$,
$D(\lambda )=f(\lambda )K(\lambda )_1^2$ satisfy relation (\ref{rk}).
Substitute these results into relations (\ref{rk1},\ref{rk2}), and after
some tedious calculations, we find that
if we take $\xi \rightarrow -i\infty$, and
$f(\lambda )=-1/e^{2i\lambda }sin(\lambda -c-\eta -s\eta )
sin(\lambda -c+s\eta)$, all relations obtained from the reflection
equation can be satisfied. So, we finally find the higher spin
reflecting matrix as
\begin{eqnarray}
A(\lambda )&=&g(\lambda )\left( e^{-4i\lambda }sin(\lambda +c-s\eta )
sin(\lambda +c+\eta +s\eta )-sin(2\lambda )sin(u+c-{\bf S^z}\eta )
e^{-i(3\lambda +c+\eta +{\bf S^z}\eta )}\right),
\nonumber \\
B(\lambda )&=&g(\lambda )sin(\eta )sin(2\lambda )
e^{-i(2\lambda -c+{\bf S^z}\eta )}{\bf S^-},
\nonumber \\
C(\lambda )&=&g(\lambda )sin(\eta )sin(2\lambda )
e^{-i(2\lambda +c+{\bf S^z}\eta )}{\bf S^+},
\nonumber \\
D(\lambda )&=&g(\lambda )\left( sin(\lambda +c-s\eta )
sin(\lambda +c+\eta +s\eta )-sin(2\lambda )sin(\lambda +c+
{\bf S^z}\eta )e^{-i(\lambda -c-\eta +{\bf S^z}\eta )}\right),
\end{eqnarray}
where $g(\lambda )=1/sin(\lambda -c-\eta -s\eta )sin(\lambda -c+s\eta )$.

Next, let us consider the higher spin reflecting matrix to the
dual reflection equation (\ref{DRK}). We suppose $K^+$ has
the similar form as $K$. By direct calculation, we can find
$R^{st_1st2}_{12}(\lambda )=I_1R_{21}(\lambda )I_1$ with $I=diag(-1,-1,1)$.
For the form (\ref{K}), we have $IK(\lambda )I=K(\lambda )$. Then with
the help of property $[M_1M_2, R(\lambda )]=0$, we can write the dual
reflection equation as
\begin{eqnarray}
R_{12}(\mu -\lambda ){K_1^+}^{st_1}(\lambda )M_1^{-1}
R_{21}(\eta -\lambda -\mu )
{K_2^+}^{st_2}(\mu )M_2^{-1}=
\nonumber \\
{K_2^+}^{st_2}(\mu )M_2^{-1}R_{12}(\eta -\lambda -\mu )
{K_1^+}^{st_1}(\lambda )M_1^{-1}
R_{21}(\mu -\lambda ).
\label{DRK1}
\end{eqnarray}
We see that there is an isomorphism between $K$ and $K^+$:
\begin{eqnarray}
K(\lambda ):\rightarrow {K^+}^{st}(\lambda )=K({\eta \over 2}-\lambda )M.
\label {ISOM}
\end{eqnarray}
Given a solution to the reflection equation (\ref{RK}), we can also
find a solution to the dual reflection equation (\ref{DRK1}).
Remark that in the
sense of the transfer matrix, the reflection equation and the dual
reflection equation are independent of each other. We can write
the higher spin reflecting matrix to the dual reflection equation as
\begin{eqnarray}
K^+(\lambda )=\left( \begin{array}{ccc}
A^+(\lambda ) &B^+(\lambda ) &0\\
B^+(\lambda ) &C^+(\lambda ) &0\\
0&0&1\end{array}\right),
\end{eqnarray}
with
\begin{eqnarray}
A^+(\lambda )&=&g^+(\lambda )
[e^{4i\lambda }sin(\lambda +\tilde {c}-\eta +\tilde {s}\eta )
sin(\lambda +\tilde {c}-2\eta -\tilde {s}\eta ).
\nonumber \\
&&-sin(2\lambda -\eta )
sin(u+\tilde {c}-\eta -\tilde {\bf S^z}\eta )
e^{i(3\lambda +\tilde {c}-\eta +\tilde {\bf S^z}\eta )}],
\nonumber \\
B^+(\lambda )&=&-g^+(\lambda )
sin(\eta )sin(2\lambda -\eta )
e^{i(2\lambda +\tilde {c}+{\eta \over 2}+\tilde {\bf S^z}\eta )}
\tilde {\bf S^-},
\nonumber \\
C^+(\lambda )&=&-g^+(\lambda )
sin(\eta )sin(2\lambda -\eta )
e^{i(2\lambda -\tilde {c}-{\eta \over 2}+\tilde {\bf S^z}\eta )}
\tilde {\bf S^+},
\nonumber \\
D^+(\lambda )&=&g^+(\lambda )[sin(\lambda +\tilde {c}-\eta +\tilde {s}\eta )
sin(\lambda +\tilde {c}-2\eta -\tilde {s}\eta )
\nonumber \\
&&-sin(2\lambda -\eta )sin(\lambda +\tilde {c}-\eta +
\tilde {\bf S^z}\eta )e^{i(\lambda -\tilde {c}+\eta
+\tilde {\bf S^z}\eta )}],
\end{eqnarray}
where $g^+(\lambda )=1/sin(\lambda -\tilde {c}+\eta +\tilde {s}\eta )
sin(\lambda -\tilde {c}-\tilde {s}\eta )$.

Thus we find the higher spin reflecting matrices for the generalized
supersymmetric $t-J$ model. We should remark that these higher spin
reflecting matrices are the kind of `singular' matrices. It can not
be constructed directly by the Sklyanin's `dressing' procedure.
In the rational limit, it reduces
to the result obtained in \cite{ZGLG}. The rational higher spin
K-matrix has been analyzied in detail by the
projecting method \cite{FS}. Our result should also be obtained
by the projecting method.

\section{Algebraic Bethe ansatz method for the generalized supersymmetric
$t-J$ model with higher spin impurities}
\subsection{First level algebraic Bethe ansatz}
We denote the double-row monodromy matrix as
\begin{eqnarray}
{\cal {T}}(\lambda )
&=&\left( \begin{array}{ccc}
{\cal {A}}_{11}(\lambda )&{\cal {A}}_{12}(\lambda ) &
{\cal {B}}_1(\lambda )\\
{\cal {A}}_{21}(\lambda )&{\cal {A}}_{22}(\lambda )
&{\cal {B}}_2(\lambda )\\
{\cal {C}}_1(\lambda )&{\cal {C}}_2(\lambda )&{\cal {D}}
(\lambda )\end{array}
\right).
\end{eqnarray}
For later discussions, we introduce the following transformations
\begin{eqnarray}
{\cal {A}}_{ab}(\lambda )=\tilde {\cal {A}}_{ab}(\lambda )
+\delta _{ab}
\frac {e^{-2i\lambda }sin(\eta )}{sin(2\lambda +\eta )}
{\cal {D}}(\lambda ).
\end{eqnarray}
As mentioned in section 2, the double-row monodromy matrix satisfies
the reflection equation (\ref{RT}), we have the following
commutation relations:
\begin{eqnarray}
{\cal {C}}_{d_1}(\lambda ){\cal {C}}_{d_2}(\mu )
&=&-\frac {\hat {r}_{12}(\lambda -\mu )_{c_2c_1}^{d_2d_1}}
{sin(\lambda -\mu +\eta )}
{\cal {C}}_{c_2}(\mu ){\cal {C}}_{c_1}(\lambda ),
\end{eqnarray}

\begin{eqnarray}
{\cal {D}}(\lambda ){\cal {C}}_{d}(\mu )
&=&\frac {sin(\lambda +\mu )sin(\lambda -\mu -\eta )}
{sin(\lambda +\mu +\eta )sin(\lambda -\mu )}
{\cal {C}}_{d}(\mu )
{\cal {D}}(\lambda )
\nonumber \\
&+&
\frac {sin(2\mu )sin(\eta )e^{i(\lambda -\mu )}}
{sin(\lambda -\mu )sin(2\mu +\eta )}  
{\cal {C}}_{d}(\lambda )
{\cal {D}}(\mu )
-\frac {sin(\eta )e^{i(\lambda +\mu )}}
{sin(\lambda +\mu +\eta )}
{\cal {C}}_b(\lambda )\tilde {\cal {A}}_{bd}(\mu ),
\end{eqnarray}

\begin{eqnarray}
\tilde {\cal {A}}_{a_1d_1}(\lambda ){\cal {C}}_{d_2}(\mu )
&=&\frac {\hat {r}_{12}(\lambda +\mu +\eta )_{a_1c_2}^{c_1b_2}
\hat {r}_{21}(\lambda -\mu )_{b_1b_2}^{d_1d_2}}
{sin(\lambda +\mu +\eta )sin(\lambda -\mu )}                
{\cal {C}}_{c_2}(\mu )
\tilde {\cal {A}}_{c_1b_1}(\lambda )
\nonumber \\
&+&
\frac {sin(\eta )e^{-i(\lambda -\mu )}}
{sin(\lambda -\mu )sin(2\lambda +\eta )}  
\hat {r}_{12}(2\lambda +\eta )_{a_1b_1}^{b_2d_1}
{\cal {C}}_{b_1}(\lambda )
\tilde {\cal {A}}_{b_2d_2}(\mu )
\nonumber \\                                                
&-&\frac {sin(2\mu )sin(\eta )e^{-i(\lambda +\mu )}}
{sin(\lambda +\mu +\eta )sin(2\lambda +\eta )sin(2\mu +\eta )}
\hat {r}_{12}(2\lambda +\eta )_{a_1b_2}^{d_2d_1}
{\cal {C}}_{b_2}(\lambda ){\cal {D}}(\mu ).
\label{AC}
\end{eqnarray}
Here the indices take values 1,2, and the matrix $\hat {r}$
is defined in (\ref{r}).

We define a reference state in the $n$-th quantum space
as $|0>_n=(0, 0, 1)^t$, and reference states for the
boundary opeators as ${\bf S^-}|0>_r=0, {\bf S^z}|0>_r=-s|0>_r,
{\bf S^+}|0>_r\not= 0$, and
$\tilde {\bf S^-}|0>_l=0, \tilde {\bf S^z}|0>_l=-\tilde {s}|0>_l,
\tilde {\bf S^+}|0>_l\not= 0$. The vacuum state is then defined as
$|0>=|0>_l\otimes_{k=1}^N|0>_k\otimes |0>_r$. Acting the double-row
monodromy matrix on this vacuum state, we have
\begin{eqnarray}                         
{\cal {B}}_a(\lambda )|0>&=&0,\nonumber \\
{\cal {C}}_a(\lambda )|0>&\not= &0,\nonumber \\
{\cal {D}}(\lambda )|0>&=&sin^{2N}(\lambda +\eta )|0>,\nonumber \\
\tilde {\cal {A}}_{ab}(\lambda )|0>&=&sin^{2N}(\lambda )
[K(\lambda )_a^b-
\delta _{ab}\frac {sin(\eta )e^{-2i\lambda }}{sin(2\lambda +\eta )}]
|0>
=W_{ab}(\lambda )sin^{2N}(\lambda )|0>,
\end{eqnarray}
where
\begin{eqnarray}
&&W_{12}(\lambda )=0, ~~~~
W_{21}(\lambda )=C(\lambda ),\nonumber \\
&&W_{11}(\lambda )=g(\lambda )\frac {e^{i\eta }sin(2\lambda )}
{sin(2\lambda +\eta )}
[e^{-i(4\lambda +2\eta )}sin(\lambda +c+e-s\eta )
sin(\lambda +c+2\eta +s\eta )
\nonumber \\
&&~~~~~~-sin(2\lambda +\eta )
sin(\lambda +c+\eta +s\eta )e^{-i(3\lambda +c+3\eta -s\eta )}],
\nonumber \\
&&W_{22}(\lambda )=-e^{-2i\lambda }\frac {sin(2\lambda )
sin(\lambda +c+\eta -s\eta )}{sin(2\lambda +\eta )sin(\lambda -c+s\eta )}.
\label{W}
\end{eqnarray}

The transfer matrix (\ref{tran}) can be written as
\begin{eqnarray}
t(\lambda )&=&-K^{+}(\lambda )_b^a{\cal {A}}_{ab}(\lambda )
+{\cal {D}}(\lambda )
\nonumber \\
&=&-K^+(\lambda )_a^b\tilde {\cal {A}}_{ba}(\lambda )
+\left( 1-\frac {sin(\eta )e^{-2i\lambda }}{sin(2\lambda +\eta )}
[A^+(\lambda )+D^+(\lambda )]\right) {\cal {D}}(\lambda ).
\label{K1}
\end{eqnarray}
Acting this transfer matrix on the ansatz of the eigenvector 
\begin{eqnarray}
{\cal {C}}_{d_1}(\mu _1)
{\cal {C}}_{d_2}(\mu _2)\cdots {\cal {C}}_{d_n}(\mu _n)|0>F^{d_1\cdots d_n},
\end{eqnarray}
where $F^{d_1\cdots d_n}$ is a function of the spectral parameters $\mu _j$,
we have
\begin{eqnarray}
&&t(\lambda )
{\cal {C}}_{d_1}(\mu _1)
{\cal {C}}_{d_2}(\mu _2)\cdots {\cal {C}}_{d_n}(\mu _n)|0>F^{d_1\cdots d_n}
\nonumber \\
&=&\frac {sin(2\lambda -\eta )sin(\lambda -\tilde {c}+\eta -\tilde {s}\eta )
sin(\lambda -\tilde {c}+2\eta +\tilde {s}\eta )}
{sin(2\lambda +\eta )sin(\lambda -\tilde {c}+\eta +\tilde {s}\eta )
sin(\lambda -\tilde {c}-\tilde {s}\eta )}
\nonumber \\
&&\times sin^{2N}(\lambda +\eta )
\prod _{i=1}^n
\frac {sin(\lambda +\mu _i)sin(\lambda -\mu _i-\eta )}
{sin(\lambda +\mu _i+\eta )sin(\lambda -\mu _i)}
{\cal {C}}_{d_1}(\mu _1)
\cdots {\cal {C}}_{d_n}(\mu _n)|0>F^{d_1\cdots d_n}
\nonumber \\
&&+sin^{2N}(\lambda )
\prod _{i=1}^n
\frac {1}
{sin(\lambda -\mu _i)sin(\lambda +\mu _i+\eta )}
{\cal {C}}_{c_1}(\mu _1)
\cdots {\cal {C}}_{c_n}(\mu _n)
t^{(1)}(\lambda )^{c_1\cdots c_n}_{d_1\cdots d_n}|0>
F^{d_1\cdots d_n}
\nonumber \\
&&+u.t.,
\end{eqnarray}
where $u.t.$ means the unwanted terms, and
$t^{(1)}(\lambda )$ is the so-called nested transfer matrix which
can be defined,
with the help of the relation (\ref{AC}), as
\begin{eqnarray}
t^{(1)}(\lambda )^{c_1\cdots c_n}_{d_1\cdots d_n}
&=&-K^+(\lambda )_b^a
\left\{ \hat {r}(\lambda +\mu _1+\eta )_{ac_1}^{a_1e_1}
\hat {r}(\lambda +\mu _2+\eta )_{a_1c_2}^{a_2e_2}\cdots
\hat {r}(\lambda +\mu _1+\eta )_{a_{n-1}c_n}^{a_ne_n}\right\}
\nonumber \\
&&\times W_{a_nb_n}(\lambda )
\left\{ \hat {r}_{21}(\lambda -\mu _n)_{b_ne_n}^{b_{n-1}d_n}
\cdots
\hat {r}_{21}(\lambda -\mu _2)_{b_2e_2}^{b_1d_2}
\hat {r}_{21}(\lambda -\mu _1)_{b_1e_1}^{bd_1}\right\} .
\end{eqnarray}
We find that this nested transfer matrix can be regarded as
a transfer matrix with reflecting boundary conditions corresponding
to the anisotropic case
\begin{eqnarray}
t^{(1)}(\lambda )=str{K^{(1)}}^+(\lambda ')
T^{(1)}(\lambda ', \{ \mu '_i\} )
K^{(1)}(\lambda ')
{T^{(1)}}^{-1}(-\lambda ', 
\{ \mu '_i )
\end{eqnarray}
with the grading $\epsilon _1=\epsilon _2=1$. Here, we denote
$\lambda '=\lambda +{\eta \over 2}, \mu '=\mu +{\eta \over 2}$.
The reflecting matrix can also be interpreted as an operator
matrix with higher spin. Explicitly, with the help of (\ref{W},\ref{K1}),
we have
\begin{eqnarray}
K^{(1)}(\lambda ')&=&e^{i\eta }\frac {sin(2\lambda '-\eta )}
{sin(2\lambda ')}\left( \begin{array}{cc}
A(\lambda ',c') &B(\lambda ',c')\\
C(\lambda ',c') &D(\lambda ',c')
\end{array}\right),
\nonumber \\
{K^{(1)}}^+(\lambda ')&=&\left( \begin{array}{cc}
A^+(\lambda '-{\eta \over 2})&B^+(\lambda '-{\eta \over 2})\\
C^+(\lambda '-{\eta \over 2})&D^+(\lambda '-{\eta \over 2})
\end{array}\right),
\end{eqnarray}
where $c'=c+{\eta \over 2}$. Note that the solution of the reflection
equation can be changed by a gauge transformation.
In order to prove that the above
defined nested transfer matrix is still a transfer matrix with
higher spin reflecting matrix, we should prove that
$K^{(1)}(\lambda ')$ and ${K^{(1)}}^+(\lambda ')$ satisfy the
reduced reflection equation and its corresponding dual reflection
equation. Indeed, it can be shown that the following reflection
equation holds,
\begin{eqnarray}
\hat {r}_{12}(\lambda '-\mu ')K^{(1)}_1(\lambda ')
\hat {r}_{21}(\lambda '+\mu ')K^{(1)}_2(\mu ')
=K^{(1)}_2(\mu ')\hat {r}_{12}(\lambda '+\mu ')
K^{(1)}_1(\lambda ')\hat {r}_{21}(\lambda '-\mu ').
\label{rk3}
\end{eqnarray}
With $M^{(1)}=diag.(e^{2i\eta },1)$, and the isomorphism
(\ref{ISOM}), we find that ${K^{(1)}}^+$ satisfies the following
relation
\begin{eqnarray}
&&\hat {r}_{12}(-\lambda '+\mu '){K^{(1)}}^+_1(\lambda ')^{st_1}
{M^{(1)}}^{-1}_1
\hat {r}_{21}(2\eta -\lambda '-\mu '){K^{(1)}}^+_2(\mu ')^{st_2}
{M^{(1)}}^{-1}_2
\nonumber \\
&&={K^{(1)}}^+_2(\mu ')^{st_2}{M^{(1)}}^{-1}_2
\hat {r}_{12}(2\eta -\lambda '-\mu ')
K^{(1)}_1(\lambda '){M^{(1)}}^{-1}_1\hat {r}_{21}(-\lambda '+\mu ').
\end{eqnarray}
By use of the cross-unitarity relation
$\hat {r}_{12}^{st_1}(2\eta -\lambda )M^{(1)}_1
\hat {r}_{21}^{st_1}(\lambda )
{M^{(1)}_1}^{-1}=sin(\lambda )sin(2\eta -\lambda )\cdot id.$,
the above relation is just the dual reflection equation which we need.

The row-to-row monodromy matrix
$T^{(1)}(\lambda ', \{ \mu '_i\} )$ (corresponding
to the periodic boundary condition) and its inverse are defined as
\begin{eqnarray}
T^{(1)}_{aa_n}(\lambda ', 
\{ \mu '_i\} )_{c_1\cdots c_n}^{e_1\cdots e_n}
&=&
\hat {r}(\lambda '+\mu '_1)_{ac_1}^{a_1e_1}
\hat {r}(\lambda '+\mu '_2)_{a_1c_2}^{a_2e_2}\cdots
\hat {r}(\lambda '+\mu '_1)_{a_{n-1}c_n}^{a_ne_n}
\\
{T^{(1)}}^{-1}_{b_na}(-\lambda ',
\{ \mu '_i\} )_{e_n\cdots e_1}^{d_n\cdots d_1}
&=&
\hat {r}_{21}(\lambda '-\mu '_n)_{b_ne_n}^{b_{n-1}d_n}
\cdots
\hat {r}_{21}(\lambda '-\mu '_2)_{b_2e_2}^{b_1d_2}
\hat {r}_{21}(\lambda '-\mu '_1)_{b_1e_1}^{ad_1}.
\end{eqnarray}
We show that a problem to find the eigenvalue 
of the original transfer matrix $t(\lambda )$
reduces to a problem to find the eigenvalue of the nested transfer
matrix $t^{(1)}(\lambda )$.
The nested transfer matrix is still a boundary case
with higher spin reflecting matrix.

In order to ensure the assumed
eigenvector is indeed the eigenvector of the transfer matrix,
$\mu _1, \cdots, \mu _n$ should satisfy the following
Bethe ansatz equations,
\begin{eqnarray}
&&\frac {sin(2\mu _j-\eta )sin(\mu _j-\tilde {c}+\eta -\tilde {s}\eta )
sin(\mu _j-\tilde {c}+2\eta +\tilde {s}\eta )}
{sin(2\mu _j+\eta )sin(\mu _j-\tilde {c}+\eta +\tilde {s}\eta )
sin(\mu _j-\tilde {c}-\tilde {s}\eta )}
sin^{2N}(\mu _j +\eta )
\nonumber \\
&&\times \prod _{i=1}^n
sin(\mu _j +\mu _i)sin(\mu _j-\mu _i-\eta )
=-{sin^{2N}(\mu _j)}
\Lambda ^{(1)}({\mu }_j),
~~~~~j=1, 2,\cdots ,n.
\end{eqnarray}
Here we have used the notation $\Lambda ^{(1)}(\lambda )$
to denote the eigenvalue
of the nested transfer matrix $t^{(1)}(\lambda )$.          

\subsection{Bethe ansatz for the six-vertex model with higher
spin reflecting matrices}
We repeat almost the same procedure as that of the first level
algebraic Bethe ansatz method. We only write down some results
without the detailed
calculations here. We have
\begin{eqnarray} 
&&e^{i\eta }\frac {sin(2\lambda '-\eta )}{sin(2\lambda ')}
D(\lambda ',c')|0>_r=-e^{-i2\lambda }
\frac {sin(2\lambda )sin(\lambda +c+\eta -s\eta )}
{sin(2\lambda +\eta )sin(\lambda -c+s\eta )}|0>_r
\equiv U_2|0>,
\\ \nonumber \\ 
&&e^{i\eta }\frac {sin(2\lambda '-\eta )}{sin(2\lambda ')}
[A(\lambda ',c')+D(\lambda ',c')\frac {sin(\eta )e^{-i2\lambda '}}
{sin(2\lambda '-\eta )}]|0>_r,
\\
&&=-e^{-i(2\lambda +\eta )}
\frac {sin(\lambda +c+\eta +s\eta )sin(\lambda -c-\eta +s\eta )}
{sin(\lambda -c-\eta -s\eta )sin(\lambda -c+s\eta )}|0>
\equiv U_1|0>_r,
\end{eqnarray}
and
\begin{eqnarray}
&&A^+(\lambda '-{\eta \over 2})|0>_r
=-e^{i(2\lambda +\eta )}
\frac {sin(\lambda +\tilde {c}-\eta +\tilde {s}\eta )}
{sin(\lambda -\tilde {c}+\eta +\tilde {s}\eta )}|0>_r
\equiv U_1^+|0>_r,\\ \nonumber \\
&&[D^+(\lambda '-{\eta \over 2})-
A^+(\lambda '-{\eta \over 2})\frac {sin(\eta )e^{-i2\lambda '}}
{sin(2\lambda '-\eta )}]|0>_r
\nonumber \\
&&=-e^{i2\lambda }\frac {sin(2\lambda -\eta )
sin(\lambda +\tilde {c}-\eta -\tilde {s}\eta )
sin(\lambda -\tilde {c}+\eta -\tilde {s}\eta )}
{sin(2\lambda )sin(\lambda -\tilde {c}-\tilde {s}\eta )
sin(\lambda -\tilde {c}+\eta +\tilde {s}\eta )}|0>_r
\equiv U_2^+|0>_r.
\end{eqnarray}
We write the nested double-row monodromy matrix as 
\begin{eqnarray}
{\cal {T}}^{(1)}(\lambda ,\{ \mu _i\} )
&=&\left( \begin{array}{cc}
{\cal {A}}^{(1)}(\lambda ) &{\cal {B}}^{(1)}(\lambda ) \\
{\cal {C}}^{(1)}(\lambda ) &{\cal {D}}^{(1)}(\lambda ) \end{array}.
\right)  
\end{eqnarray}
From the results obtained above, we know that this double-row monodromy
matrix also satisfies the reflection equation (\ref{rk3}). Considering
the transformation
\begin{eqnarray}
{\cal {A}}^{(1)}(\lambda )=\tilde {\cal {A}}^{(1)}(\lambda )
-\frac {sin(\eta )e^{-2i\lambda }}{sin(2\lambda -\eta )}
{\cal {D}}^{(1)}(\lambda ),
\end{eqnarray}
we can find the following commutation relations which are useful
for the algebraic Bethe ansatz method,
\begin{eqnarray}
{\cal {D}}^{(1)}(\lambda )
{\cal {C}}^{(1)}(\mu )
&=&\frac {sin(\lambda -\mu +\eta )sin(\lambda +\mu )}
{sin(\lambda -\mu )sin(\lambda +\mu -\eta )}
{\cal {C}}^{(1)}(\mu )
{\cal {D}}^{(1)}(\lambda )
\nonumber \\
&-&\frac {sin(2\mu )sin(\eta )e^{i(\lambda -\mu )}}
{sin(\lambda -\mu )sin(2\mu -\eta )}
{\cal {C}}^{(1)}(\lambda )
{\cal {D}}^{(1)}(\mu )
+\frac {sin(\eta )e^{i(\lambda +\mu )}}{sin(\lambda +\mu -\eta )}
{\cal {C}}^{(1)}(\lambda )
\tilde {\cal {A}}^{(1)}(\mu ),
\end{eqnarray}

\begin{eqnarray}
\tilde {\cal {A}}^{(1)}(\lambda )
{\cal {C}}^{(1)}(\mu )
&=&\frac {sin(\lambda -\mu -\eta )sin(\lambda +\mu -2\eta )}
{sin(\lambda -\mu )sin(\lambda +\mu -\eta )}
{\cal {C}}^{(1)}(\mu )
\tilde {\cal {A}}^{(1)}(\lambda )
\nonumber \\
&+&\frac {sin(\eta )sin(2\lambda -2\eta )
e^{-i(\lambda -\mu )}}
{sin(\lambda -\mu )sin(2\lambda -\eta )}
{\cal {C}}^{(1)}(\lambda )
\tilde {\cal {A}}^{(1)}(\mu )
\nonumber \\
&&-\frac {sin(2\mu )sin(2\lambda -2\eta )sin(\eta )e^{-i(\lambda +\mu )}}
{sin(\lambda +\mu -\eta )sin(2\lambda -\eta )sin(2\mu -\eta )}
{\cal {C}}^{(1)}(\lambda )
{\cal {D}}^{(1)}(\mu ),
\end{eqnarray}
\begin{eqnarray}
{\cal {C}}^{(1)}(\lambda )
{\cal {C}}^{(1)}(\mu )
&=&
{\cal {C}}^{(1)}(\mu )
{\cal {C}}^{(1)}(\lambda  ).
\end{eqnarray}
We thus finally obtain the eigenvalues of the nested transfer matrix
as
\begin{eqnarray}
\Lambda ^{(1)}(\lambda ')
&=&-U_1^+U_1
\prod _{i=1}^n[sin(\lambda '+\mu '_i)
sin(\lambda '-\mu '_i)]
\prod _{l=1}^m \left\{
\frac {sin(\lambda '-{\mu '}^{(1)}_l-\eta )
sin(\lambda '+{\mu '}^{(1)}_l-2\eta )}
{sin(\lambda '-{\mu '}^{(1)}_l)
sin(\lambda '+{\mu '}^{(1)}_l-\eta )}\right\} 
\nonumber\\ 
&&-U_2^+U_2
\prod _{i=1}^n
[sin(\lambda '+{\mu '}_i-\eta )
sin(\lambda '-{\mu '}_i-\eta )] 
\nonumber \\
&&\prod _{l=1}^m
\left\{ \frac {sin(\lambda '-{\mu '}^{(1)}_l+\eta )
sin(\lambda '+{\mu '}^{(1)}_l)}
{sin(\lambda '-{\mu '}^{(1)}_l)
sin(\lambda '+{\mu '}^{(1)}_l-\eta )}\right\} ,
\end{eqnarray}
where ${\mu '}^{(1)}_1, \cdots ,{\mu '}^{(1)}_m$
should satisfy the corresponding Bethe ansatz equations.
In waht follows, we give a summary of our main result.

\subsection{Result}
The eigenvalues of the transfer matrix for the generalized supersymmetric
$t-J$ model are given as follows:
\begin{eqnarray}
\Lambda (\lambda )
&=&\frac {sin(2\lambda -\eta )sin(\lambda -\tilde {c}+\eta -\tilde {s}\eta )
sin(\lambda -\tilde {c}+2\eta +\tilde {s}\eta )}
{sin(2\lambda +\eta )sin(\lambda -\tilde {c}+\eta +\tilde {s}\eta )
sin(\lambda -\tilde {c}-\tilde {s}\eta )}
\nonumber \\
&&\times sin^{2N}(\lambda +\eta )
\prod _{i=1}^n
\frac {sin(\lambda +\mu _i)sin(\lambda -\mu _i-\eta )}
{sin(\lambda +\mu _i+\eta )sin(\lambda -\mu _i)}
\nonumber \\
&&+sin^{2N}(\lambda )
\prod _{i=1}^n
\frac {1}
{sin(\lambda -\mu _i)sin(\lambda +\mu _i+\eta )}
\Lambda ^{(1)}(\lambda ),
\end{eqnarray}

\begin{eqnarray}
\Lambda ^{(1)}(\lambda )
&=&-
\frac {sin(\lambda +c+\eta +s\eta )sin(\lambda -c-\eta +s\eta )
sin(\lambda +\tilde {c}-\eta +\tilde {s}\eta )}
{sin(\lambda -c-\eta -s\eta )sin(\lambda -c+s\eta )
sin(\lambda -\tilde {c}+\eta +\tilde {s}\eta )}
\nonumber \\
&&\times \prod _{i=1}^n[sin(\lambda +\mu _i+\eta )
sin(\lambda -\mu _i)]
\prod _{l=1}^m \left\{
\frac {sin(\lambda -\mu ^{(1)}_l-\eta )
sin(\lambda +\mu ^{(1)}_l-\eta )}
{sin(\lambda -\mu ^{(1)}_l)
sin(\lambda +\mu ^{(1)}_l)}\right\} 
\nonumber\\ 
&&-
\frac {sin(2\lambda -\eta )
sin(\lambda +\tilde {c}-\eta -\tilde {s}\eta )
sin(\lambda -\tilde {c}+\eta -\tilde {s}\eta )
sin(\lambda +c+\eta -s\eta )}
{sin(2\lambda +\eta )sin(\lambda -\tilde {c}-\tilde {s}\eta )
sin(\lambda -\tilde {c}+\eta +\tilde {s}\eta )
sin(\lambda -c+s\eta )}
\nonumber \\
&&\times \prod _{i=1}^n
[sin(\lambda +\mu _i)
sin(\lambda -\mu _i-\eta )] 
\prod _{l=1}^m
\left\{ \frac {sin(\lambda -\mu ^{(1)}_l+\eta )
sin(\lambda +\mu ^{(1)}_l+\eta )}
{sin(\lambda -\mu ^{(1)}_l)
sin(\lambda +\mu ^{(1)}_l)}\right\} ,
\end{eqnarray}
where $\mu _1, \cdots ,\mu _n$ and $\mu ^{(1)}_1, \cdots ,\mu ^{(1)}_m$
should satisfy the Bethe ansatz equations

\begin{eqnarray}
&&
\frac {sin(\mu _j^{(1)}+c+\eta +s\eta )sin(\mu _j^{(1)}-c-\eta +s\eta )
sin(\mu _j^{(1)}+\tilde {c}-\eta +\tilde {s}\eta )
sin(\mu _j^{(1)}-\tilde {c}-\tilde {s}\eta )               
}
{sin(\mu _j^{(1)}-c-\eta -s\eta )sin(\mu _j^{(1)}+c+\eta -s\eta )
sin(\mu _j^{(1)}+\tilde {c}-\eta -\tilde {s}\eta )
sin(\mu _j^{(1)}-\tilde {c}+\eta -\tilde {s}\eta )
}
\nonumber \\
&&
=
\prod _{i=1}^n
\frac {sin(\mu _j^{(1)}+\mu _i)
sin(\mu _j^{(1)}-\mu _i-\eta )} 
{sin(\mu _j^{(1)}+\mu _i+\eta )
sin(\mu _j^{(1)}-\mu _i)}
\prod _{l=1, \not=j}^m
\frac {sin(\mu _j^{(1)}-\mu ^{(1)}_l+\eta )
sin(\mu _j^{(1)}+\mu ^{(1)}_l+\eta )}
{sin(\mu _j^{(1)}-\mu ^{(1)}_l-\eta )
sin(\mu _j^{(1)}+\mu ^{(1)}_l-\eta )},
\nonumber \\
&&\hskip 10truecm~~~~j=1, \cdots , m,
\end{eqnarray}
and 
\begin{eqnarray}
&&\frac 
{sin(\mu _j+\tilde {c}-\eta -\tilde {s}\eta )
sin(\lambda +c+\eta -s\eta )
}
{sin(\mu _j-\tilde {c}+2\eta +\tilde {s}\eta )
sin(\lambda -c+s\eta )}
=
\frac {sin^{2N}(\mu _j+\eta )}{sin^{2N}(\mu _j)}
\prod _{l=1}^m
\frac {sin(\mu _j-\mu _l^{(1)})
sin(\mu _j+\mu _l^{(1)})}
{sin(\mu _j-\mu _l^{(1)}+\eta )
sin(\mu _j+\mu _l^{(1)}+\eta )},
\nonumber \\
&&\hskip 10truecm~~~~j=1,\cdots ,n.
\end{eqnarray}

\section{Summary}
In this paper, we have studied the generalized supersymmetric $t-J$ model
with Kondo impurities in the boundaries. Using
the higher spin L operator of XXZ
Heisenberg chain and the general diagonal solution to the reflection
equation for six vertex model, we find a higher spin reflecting
matrix for the generalized supersymmetric $t-J$ model. Applying
the graded algebraic Bethe ansatz method, we obtain the eigenvalues
of the transfer matrix for the $t-J$ model with higher spin boundaries.

It is interesting to solve this problem in other background gradings,
for example, FBF or BFF. The higher spin reflecting matrix
should be constructed from the BF or FB six vertex models.
The analysis of ground state properties, low-lying excitations and
thermodynamic Bethe ansatz
is always worth performing. 

One can find that the $SU_q(2)$ higher spin reflecting matrix also
satisfy the reflection equation of $SU_q(N)$ model. The eigenvalues
of $SU_q(N)$ model with $SU_q(2)$ higher spin boundary impurities can be
obtained by using the nested algebraic Bethe ansatz method. Actually,
the $SU_q(2)$ higher spin boundary impurities could be embeded into
$SU_q(M|N)$ spin chains with $M\ge= 2$ or $N\ge 2$.

After we put our article to the cond-mat e-print archive,
X.Y.Ge and H.Q.Zhou
inform us that they solve the same problem independently\cite{GGLZ}.

\vskip 1.5truecm
\noindent {\bf Acknowlegements}:
One of the authors, H.F. is supported by the Japan Society for
the Promotion of Science. He would like to thank the hospitality
of Department of Physics, University of Tokyo and the help of 
Wadati's group. We thank X.Y.Ge, H.Q.Zhou and K.L.Hur for valuable
suggestions and comments.
We thank B.Y.Hou, K.J.Shi, W.L.Yang and L.Zhao
for useful discussions.
This work was partly supported by
the NSFC, Clibming project and NWU Teachers' fund.

\end{document}